\documentclass[12pt]{article}

\usepackage{amstext, graphicx, amsmath, latexsym, amssymb, amscd, amsfonts, float,placeins,color}
\usepackage{multirow}
\usepackage{booktabs}
\usepackage{bbm}
\usepackage{harpoon}
\usepackage{amsthm}
\usepackage{tikz}
\usepackage{siunitx}
\usepackage{amsmath,bm}
\usepackage[title]{appendix}
\usepackage[noend]{algpseudocode}
\usepackage{algorithmicx,algorithm}
\usepackage[colorlinks,linkcolor=blue]{hyperref}

\newtheorem{Theorem}{\bf Theorem}[section]		
\newtheorem{Lemma}{\bf Lemma}[section]
\newtheorem{Proposition}{\bf Proposition}[section]
\newtheorem{Definition}{\bf Definition}[section]
\newtheorem{Example}{\bf Example}[section]
\newtheorem{Remark}{\bf Remark}[section]

\numberwithin{equation}{section}
\numberwithin{figure}{section}
\allowdisplaybreaks[4]

\begin{document}
	
	\title{Systemic risk measures with market volatility
		\thanks{The research of Fei Sun is supported bysupported by the National Natural Science Foundation of China (12401620) and the Special Foundation in Key Fields for Universities of Guangdong Province (2023ZDZX4060). The research of Jieming Zhou is supported by the Natural Science Foundation of Hunan Province (2023JJ30381), the Changsha Municipal Natural Science Foundation (kq2208159).}}

	\author{Fei Sun$^{1}$, Jieming Zhou$^{2}$\footnote{Corresponding Author: Jieming Zhou,  E-mail: jmzhou@hunnu.edu.cn} }

	\date{}
	\maketitle

	\begin{center}\small{$^1$ School of Mathematics and Computational Science, Wuyi University, Jiangmen, 529020, P.R. China}
\end{center}
\begin{center} \small {$^2$ MOE-LCSM, School of Mathematics and Statistics, Hunan Normal University, Changsha, 410081, P.R. China} 
\end{center}


	\begin{abstract}
		\noindent  
		Systemic risk measures are crucial for the stability of financial markets, yet classical formulations fail to capture the complexity of market volatility. We propose a new framework for systemic risk measurement on the variable-exponent Bochner-Lebesgue space  $L^{p(\cdot)}$, where the exponent $p(\cdot)$ is a random variable rather than a deterministic constant parameter, thereby inherently encoding latent market volatility. By constructing suitable deterministic auxiliary functions and single-firm risk measures, we decompose the quantification of systemic risk in $L^{p(\cdot)}$ into two sequential steps, ultimately deriving its dual representations. Several examples are provided to illustrate the theoretical results.

		\medskip
		
	\end{abstract}
	\thispagestyle{empty}
	\newpage

	\section{Introduction}
	\label{sec:1}
	

	Systemic risk refers to the risk arising from macroeconomic shocks, policy changes, or similar aggregate disturbances whose effects spread throughout the entire market or financial system. Due to its contagious and destructive nature, systemic risk can undermine market confidence, cause sharp declines in asset prices, and ultimately trigger financial crises. As a result, systemic risk represents the greatest threat to financial stability, making its accurate measurement essential for protecting economic security and maintaining public trust. In fact, systemic risk measurement, which precisely assesses financial system vulnerability under extreme conditions, provides regulators with a quantitative basis for setting capital buffers and liquidity requirements. Additionally, these assessments enable financial institutions to optimize asset allocation and prevent excessive risk accumulation. Furthermore, such analyses help identify critical nodes within the risk network, facilitating targeted supervision. Central to this effort is the development of a robust framework for quantifying systemic risk.


	Systemic risk measures are initially formalized through an axiomatic approach by Chen et al. \cite{8}, who demonstrated that while offsetting losses on individual assets with gains elsewhere constitutes rational portfolio management, such practices are inappropriate for macroeconomic assessment. This is because heterogeneous equity ownership structures across financial institutions render the aggregate social impact non-additive. Rejecting the reductionist view of the financial system as a single composite portfolio, Chen et al. \cite{8} argued that monetary risk measures cannot be mechanically applied to economy-wide aggregates without accounting for inter-institutional interactions. Consequently, they decompose systemic risk into an aggregation functional and a scalar risk component, establishing a mechanism to attribute individual institutions' contributions.
	Subsequent developments include Kromer et al. \cite{26}, who generalized this axiomatic framework to general probability spaces. Complementing this line of research, Doldi and Frittelli \cite{27}, Kromer et al. \cite{28}, and Hoffmann et al. \cite{29} examined dynamic systemic risk measures and their associated time consistency properties from multiple theoretical perspectives. Additionally, Ararat and Rudloff \cite{30} derived dual representations for these models.
	Additional contributions to systemic risk measurement include the works of Acharya et al. \cite{1}, Armenti et al. \cite{3}, Biagini et al. \cite{6}, Brunnermeier and Cheridito \cite{7}, Feinstein et al. \cite{11}, Gauthier et al. \cite{13}, Tarashev et al. \cite{23}, and the references therein.

	Recent seismic shifts in the global economic order, including geopolitical tensions and the ongoing retreat from globalization, have significantly increased price volatility.
	However, existing systemic risk measures remain fundamentally ill-suited to volatile market conditions. Current quantification methodologies have three critical limitations. First, conventional approaches rely predominantly on historical data. Yet, markets now exhibit low-frequency and high-severity loss dynamics, causing regulators to systematically underestimate risk exposures during tail events. Second, parameter estimation lags significantly, preventing the real-time capture of market sentiment. 
	Third, elevated high-frequency market volatility distorts fundamental asset pricing mechanisms, triggering liquidity crises and simultaneously amplifying measurement errors—ultimately causing risk assessments to diverge significantly from realized losses.
	Therefore, developing a systemic risk framework specifically designed for volatile market environments has become essential to safeguarding financial stability.

	In this paper, we investigate systemic risk measures on the variable-exponent Bochner-Lebesgue space $L^{p(\cdot)}$. The framework of this study is based on the following mapping: 
	\begin{align*}
		\rho: L^{p(\cdot)}\ni f \mapsto \rho(f)\in\mathbb{R}\cup \{+\infty\}
	\end{align*}
	which quantifies the characteristics of systemic risk with market volatility.
	In fact, one of the core characteristics of financial market volatility lies in its time-varying and state-dependent nature. During tranquil periods, the distribution of asset returns is relatively concentrated, with risks primarily manifested as variations around the mean. In contrast, during periods of stress, the market experiences not only intensified fluctuations but, more critically, a shift in the shape of its distribution, particularly a pronounced emergence of tail risk, which leads to a substantial increase in the probability of extreme losses. The classical risk measurement framework, which relies on the fixed-exponent $L^{q}$ space, implicitly assumes a globally uniform moment condition. It presumes that the metric for the magnitude of risk—that is, the emphasis placed on the moments of loss—remains constant regardless of the market state. This assumption clearly fails to capture the dynamic features described above. A space with a fixed $q$ either overemphasizes the tail during calm periods (if $q$ is relatively large) or underestimates extreme losses during crises (if $q$ is relatively small).
	The study of systemic risk measures on the space $L^{p(\cdot)}$ is precisely aimed at incorporating the quality—not just the quantity—of volatility into the framework of risk measurement. Here, the variable exponent $p(\cdot)$ can be regarded as an endogenous risk-sensitivity adjuster, whose value is directly driven by the prevailing market state, such as implied volatility or the level of systemic stress. When market volatility rises and is accompanied by an accumulation of tail risk, $p(\cdot)$ increases accordingly, causing the norm of the space to penalize large losses more severely, thereby automatically enhancing the sensitivity of the risk measure to extreme events. Conversely, during periods of lower volatility and milder return distributions, $p(\cdot)$ decreases, shifting the focus of the measure toward the central tendency of losses. This design imparts adaptability to the risk measure itself. Rather than applying a fixed scale across all market regimes, it allows the measurement criterion to adjust dynamically in response to the evolving nature of risk in the market environment.

	It should be noted that the variable-exponent Bochner-Lebesgue space $L^{p(\cdot)}$ was first introduced by Orlicz \cite{19}.
	Subsequently, Cheng and Xu \cite{9} conducted a comprehensive study of $L^{p(\cdot)}$ and, for the first time, provided a characterization of its dual space.
	Further studies on $L^{p(\cdot)}$ include those by Almeida et al. \cite{2}, Diening et al. \cite{10}, Harjulehto et al. \cite{14}, H\"{a}st\"{o} \cite{15}, Kempka \cite{16,17}, Kov\'{a}\v{c}ik and R\'{a}kosn\'{\i}k \cite{18}, Xu \cite{24,25}, and the references therein.

	It turns out that constructing systemic risk measures on the space $L^{p(\cdot)}$ can be decomposed into two sequential steps.
	First, by leveraging the characteristics of market volatility, we establish a deterministic mapping that dynamically normalizes systemic risk positions. Through a standardized transformation, multi-source systemic risk is converted into a univariate measurable quantity, effectively filtering out noise from market fluctuations.
	Second, a single-firm risk measure is introduced to evaluate normalized positions. By decoupling the systemic risk structure in a staged manner, this approach retains the conventional ability to capture common market risk. Additionally, its modular design enhances cross-sector comparability and allows for dynamic recalibration, thereby improving the precision of systemic risk measurement in volatile markets.

	The principal contribution of this paper lies in the development of a novel framework for systemic risk quantification that explicitly addresses the challenges posed by volatile market conditions. The contribution comprises three key aspects, which we elaborate as follows.
	First, we establish a foundational innovation by introducing the variable-exponent Bochner-Lebesgue space $L^{p(\cdot)}$—wherein the random exponent $p(\cdot)$ encodes dynamic market volatility—as the mathematical domain for modeling systemic risk positions. This is the application of such function spaces to systemic risk analysis.
	Second, within this framework, we develop systemic risk measures that inherently incorporate market volatility dynamics. We further advance the theoretical foundations by proving a structural decomposition theorem: any systemic risk measure in $L^{p(\cdot)}$ can be decomposed into a convex deterministic component and a single-firm risk component, providing novel insights into risk composition under volatile market conditions.
	Third, we derive the dual representations of these models by applying duality theory for $L^{p(\cdot)}$. Notably, obtaining these dual representations presents particular methodological challenges, as significant theoretical gaps exist between the established characterization theory of dual spaces with variable exponents and our derived representations. To bridge these gaps, we employ a novel approach involving carefully designed construction procedures.

	The remainder of this paper is organized as follows. In Section~\ref{sec:2}, we briefly review the definition and main properties of the variable-exponent Bochner-Lebesgue space $L^{p(\cdot)}$. In Section~\ref{sec:3}, we develop the definitions of systemic risk measures in $L^{p(\cdot)}$ as well as the definitions of the convex deterministic function and single-firm risk measures. Section~\ref{sec:4} discusses the construction of the systemic risk measures in $L^{p(\cdot)}$. Section~\ref{sec:5} is devoted to the dual representations of systemic risk measures in $L^{p(\cdot)}$.
	Finally, in Section~\ref{sec:6}, examples of systemic risk measures in $L^{p(\cdot)}$ are presented.

	\section{Preliminary information}
	\label{sec:2}
	This paper is devoted to constructing systemic risk quantification models in the variable-exponent Bochner-Lebesgue space $L^{p(\cdot)}$ and deriving their dual representations. The variable-exponent Bochner-Lebesgue space was first introduced and systematically studied by Cheng and Xu \cite{9}. This space is particularly well-suited to financial markets subject to stochastic volatility because the random exponent $p(\cdot)$ is itself a random variable whose fluctuations encode the time-varying intensity of market turbulence. Interpreting $p(\cdot)$ as the instantaneous sensitivity of investments to such turbulence, the variable-exponent framework endogenously captures evolving risk exposures. In contrast, the classical $L^{q}$ space fixes $q$ as a constant, rendering it incapable of adapting to the dynamic sensitivity patterns inherent in volatile market conditions. Hence, the variable-exponent Bochner-Lebesgue space provides a more appropriate domain for modeling risk positions under market volatility.

	In this section, we briefly introduce the definition and main properties of variable-exponent Bochner-Lebesgue spaces, along with some preliminary information that will be used throughout this paper.

	Let $(\Omega,\mathcal{F},\mu)$ be a $\sigma$-finite complete measurable space, and  denote by $L^{q}:=L^{q}(\Omega,\mathcal{F},\mu)$ the linear space of equivalence classes of $\mathcal{F}$-measurable function $X:\Omega\rightarrow \mathbb{R}$ such that $\int_{\Omega}|X(\omega)|^{q}d\mu<\infty$ for $q\in [1,\infty)$ and ess.sup$_{\omega\in\Omega}|X|<\infty$ for $q=\infty$, where the $|\cdot|$ is an arbitrary fixed norm on $\mathbb{R}$. We also denote by $L^{s}$ the dual space of $L^{q}$ and $L^{q}_{+} := \{ X\in L^{q} \ \big| \ X(\omega) \geq 0 \textrm{ for all } \omega \in \Omega\}$.

	Let $E$ be a given reflexive Banach space with norm $\|\cdot\|$ and dual space $E^{\ast}$ having the Radon-Nikod\'{y}m property. We assume that $E^{\ast}$ is partially ordered by a given cone $K_{0}$ and $E$ is partially ordered by $K$, where $K:=\{x\in E \ \big| \ \langle y, x\rangle\geq0\textrm{ for any } y\in K_{0}\}$ is the positive dual cone of $K_{0}$. We also suppose that the num\'{e}raire asset is some interior point $z\in int(K)$.

	\begin{Remark}\label{R1}
		The partial order relation $\geq_{K}$ is defined as follows, for any $x,y\in E$
		\begin{displaymath}
			x\geq_{K} y\Leftrightarrow x-y\in K.
		\end{displaymath}
	\end{Remark}

	\begin{Remark}
		The cone $K$ consists of the ‘admissible’ price functionals. 
		For example, when managing with a diversified portfolio, investors do not need each individual investment to be profitable. Some investments may incur losses, but these are acceptable as long as the investor achieves overall profitability.
		$K$ is also a solvency set of financial positions and reflects how a group of investors collectively interprets the common concept of the cost of financial positions.
	\end{Remark}

	Banach space valued Bochner-Lebesgue spaces with variable-exponents were first introduced by Cheng and Xu \cite{9}. We now recall the definition and related properties of this special space. We denote the set of all $\mathcal{F}$-measurable functions $p(\cdot):\Omega\rightarrow [1,\infty]$ by $\mathcal{S}(\Omega, \mu)$; these functions are called variable-exponent functions with respect to $\Omega$. For a function $p(\cdot)\in \mathcal{S}(\Omega, \mu)$, we define $p^{\prime}(\cdot)\in \mathcal{S}(\Omega, \mu)$ by $1/p(\omega)+1/p^{\prime}(\omega)=1$.
	The following definitions and properties are taken from the work of Cheng and Xu \cite{9}.
	
	\begin{Definition}
		A function $f:\Omega\rightarrow E$ is strongly $\mathcal{F}$-measurable if there exists a sequence $\{f_{n}\}_{n\geq1}$ of $\mu$-simple functions converging to $f$ $\mu$-almost everywhere.
	\end{Definition}

	\begin{Definition}
		The variable-exponent Bochner-Lebesgue space, denoted by $L^{p(\cdot)}$, is the collection of all strongly $\mathcal{F}$-measurable functions $f:\Omega\rightarrow E$ endowed with the norm
		\begin{displaymath}
			\|f\|_{L^{p(\cdot)}}:=\inf\{\lambda>0 \ \big| \  \rho_{p(\cdot)}(f/\lambda)\leq1\}
		\end{displaymath}
		where
		\begin{displaymath}
			\rho_{p(\cdot)}(f):=\int_{\Omega}\|f(\omega)\|^{p(\omega)}d\mu(\omega)\quad \textrm{and}\quad p(\cdot)\in \mathcal{S}(\Omega, \mu).
		\end{displaymath}
	\end{Definition}
	
	The partial order  $ f\geq_{K} g $ on variable-exponent Bochner-Lebesgue space is defined as $ f(\omega)\geq_{K} g(\omega) $  for any $ \omega \in \Omega $.

	\begin{Remark}\label{L2}
		The dual of $L^{p(\cdot)}$ is characterized by the linear isomorphism $L^{p^{\prime}(\cdot)}(\Omega, E^{\ast})\ni g\mapsto V_{g}\in (L^{p(\cdot)})^{\ast}$ as follows:
		\begin{displaymath}
			\langle V_{g}, f\rangle  = \int_{\Omega}\langle g, f\rangle d\mu, \qquad \textrm{for any}\quad f\in L^{p(\cdot)}.
		\end{displaymath}
		See Theorem 2 of Cheng and Xu \cite{9}.
	\end{Remark}

	The principal contribution of Cheng and Xu \cite{9} is the characterization of the dual of the variable-exponent Bochner-Lebesgue space (see, Remark \ref{L2}) together with its associated properties. While duality is central to functional analysis, it is also fundamental in risk quantification for constructing dual representations of risk measures. Building upon this dual structure, we develop a systemic risk framework and derive explicit dual representations through novel and carefully designed construction procedures.

	\section{The definition of systemic risk measures}
	\label{sec:3}
	In most financial markets, systemic risk is defined as the risk of deterioration among institutions and other market participants in a chain-like fashion, potentially impacting the entire financial system negatively. Notably, the risk of a `domino effect' certainly seems central to the concept of systemic risk, as is the risk of a triggering event that causes the first domino to fall. Generally, systemic risk refers to the potential for significant volatility in asset prices, corporate liquidity issues, bankruptcies, and efficiency losses caused by economic shocks.
	Hence, the importance of systemic risk quantification models is self-evident. These models offer precise diagnostics of financial system fragility under extreme scenarios, providing regulators with rigorous data to calibrate capital buffers and liquidity requirements, while simultaneously enabling financial institutions to optimize asset allocation and prevent risk accumulation.

	
	In this section, we will introduce the definitions of systemic risk measures.
	Here, we adopt an axiomatic framework to define systemic risk measures in $L^{p(\cdot)}$. 
	For preparation, we define two special functions.

	\begin{Definition}\label{D31}
		Recall that $E$ denotes a reflexive Banach space (see, Section~\ref{sec:2}).
		A convex deterministic function is a mapping $\phi:E \rightarrow \mathbb{R}$ that satisfies the following properties:
		\begin{description}
			
			\item[A0] Surjectivity: $\phi(E)=\mathfrak{L}$ with $\mathfrak{L} = \mathbb{R}$ or $\mathfrak{L} = \mathbb{R}_{+}$;
			\item[A1] Monotonicity: for any $x,y\in E$, $x\geq_{K} y$ implies $\phi(x)\geq\phi(y)$;
			\item[A2] Convexity: for any $x,y\in E$ and $\lambda\in[0,1]$, $\phi(\lambda x+(1-\lambda)y)\leq \lambda\phi(x)+(1-\lambda)\phi(y)$;
			\item[A3] There exists a constant $q\in [1,\infty]$, such that
			\begin{align*}
				\big\{ f_{\phi}: \Omega \ni \omega \mapsto \phi(f(\omega)) \in \mathbb{R} \ \big| \ f \in L^{p(\cdot)} \big\} = L^{q} \textrm{ for }   \mathfrak{L} = \mathbb{R}
			\end{align*}
			\begin{align*}
				\big(\textrm{respectively} \ \big\{ f_{\phi}: \Omega \ni \omega \mapsto \phi(f(\omega)) \in \mathbb{R}_{+} \ \big| \ f \in L^{p(\cdot)} \big\} = L^{q}_{+} \  \textrm{ for }   \mathfrak{L} = \mathbb{R}_{+}\big).
			\end{align*}
		\end{description}
	\end{Definition}
	
	\begin{Remark}
		The order of  $\mathbf{A1}$ is the partial order induced by the cone $K$ as defined in Remark~\ref{R1}. Thus, the Banach space $E$ is partially ordered by the given cone $K$.  $\mathbf{A0}$ states that the function $\phi$ is non-constant and unbounded from above.
	\end{Remark}

	The following lemma provides a sufficient condition for $\phi: E \rightarrow \mathbb{R}$ to satisfy $\mathbf{A3}$ with $ \mathfrak{L} = \mathbb{R} $. 
	

	\begin{Lemma}\label{L31}
	For a $\phi: E \rightarrow \mathbb{R}$ satisfying $\mathbf{A0}$-$\mathbf{A2}$, it satisfies $\mathbf{A3}$ with $ \mathfrak{L} = \mathbb{R} $ if 
		\begin{description}
			\item[($\bullet$)]  There exists $q\geq 1$ such that
			$\|T_{\phi}(Z)\|_{q}< \infty$  and $\|T^{-1}_{\phi}(Z)\|_{q}< \infty$ for any $Z\in L^{q}$. Here, $T_{\phi}$ is defined as $T_{\phi}: \mathbb{R}\ni a \mapsto \phi(az)\in \mathbb{R}$ with $z\in int(K)$.
		\end{description}
	\end{Lemma}
	
	\noindent \textbf{Proof.}
	For any $f\in L^{p(\cdot)}$, we consider $Z$, which is defined by
	\begin{displaymath}
		Z(\omega):=\left\{ \begin{array}{ll}
			\max\{a| f(\omega)\leq_{K} az\}, & w\in A\\
			\min\{a| az\leq_{K} f(\omega)\}, & w\in \Omega\setminus A
		\end{array} \right.
	\end{displaymath}
	where $A=\{\omega \in \Omega \mid \phi(f(\omega))\geq0\}$. From  $\mathbf{A1}$, we obtain
	$
	0 \leq \phi(f(\omega)) \leq \phi(Z(\omega)z) $ for any $ w\in A
	$
	and
	$
	0 > \phi(f(\omega)) \geq \phi(Z(\omega)z) $ for any $ w\in \Omega\setminus A.
	$
	This leads to
	$
	|\phi(f(\omega))| \leq |\phi(Z(\omega)z)| = |T_{\phi}(Z(\omega))| $ for any $ w\in \Omega.
	$
	It follows that for any $f\in L^{p(\cdot)}$, there exists $Z\in L^{q}$ such that
	$
	\mathbb{E}[|\phi(f)|^{q}]\leq \mathbb{E}[|T_{\phi}(Z)|^{q}] $ for $ q<\infty
	$
	and
	$
	\inf\{b\in \mathbb{R}\mid |\phi(f)|\leq b\}\leq \inf\{b\in \mathbb{R}\mid |T_{\phi}(Z)|\leq b\} $ for $ q=\infty.
	$
	Using the assumption $(\bullet)$, as $\|T_{\phi}(Z)\|_{q}< \infty$ for any $Z\in L^{q}$, we have
	$
	\|\phi(f)\|_{q}\leq \|T_{\phi}(Z)\|_{q} < \infty $ for any  $ f\in L^{p(\cdot)}.$
	Thus, $\big\{ f_{\phi}: \Omega \ni \omega \mapsto \phi(f(\omega)) \in \mathbb{R} \ \big| \ f \in L^{p(\cdot)} \big\} \subseteq L^{q}$.
	We know that the $T_{\phi}$ is a bijective function since $\phi$ satisfies  $\mathbf{A0}$-$\mathbf{A2}$.
	Therefore, for any $X\in L^{q}$, we can define $Y$ by
	$
	Y(\omega):= T^{-1}_{\phi}(X(\omega)) $ for all $ \omega \in \Omega.
	$
	As $\|T^{-1}_{\phi}(Z)\|_{q}< \infty$ for any $Z\in L^{q}$, it is relatively simple to check that $Y \in L^{q}$. Hence, there exists a vector
	$Yz \in L^{p(\cdot)}$ such that
	$
	\phi(Yz)= T_{\phi}(Y) = X.
	$
	Thus, $L^{q} \subseteq \big\{ f_{\phi}: \Omega \ni \omega \mapsto \phi(f(\omega)) \in \mathbb{R} \ \big| \ f \in L^{p(\cdot)} \big\}$, and we arrive at $\big\{ f_{\phi}: \Omega \ni \omega \mapsto \phi(f(\omega)) \in \mathbb{R} \ \big| \ f \in L^{p(\cdot)} \big\} = L^{q}$.  \qed
	

	In fact, the convex deterministic function $\phi$ is used to transform the uncertainty of systemic risk into certainty. However, to measure systemic risk in $L^{p(\cdot)}$, we still require a  convex single-firm risk measure to quantify the risk simplified by the convex deterministic function.

	\begin{Definition}\label{D32}
		A convex single-firm risk measure on $L^{q}$ with $q\geq 1$ is a functional 
		\begin{align*}
			\varrho: L^{q}\ni f \mapsto  \varrho(f)\in \mathbb{R}\cup \{+\infty\}
		\end{align*}
		that satisfies the following properties:
		\begin{description}
			\item[B1] Monotonicity: for any $X,Y\in L^{q}$, $X\geq Y$ implies  $\varrho(X)\geq \varrho(Y)$;
			\item[B2] Convexity: for any $X,Y\in L^{q}$ and $\lambda\in[0,1]$, $\varrho\big(\lambda X+(1-\lambda)Y\big)\leq \lambda\varrho(X)+(1-\lambda)\varrho(Y)$;
			\item[B3] Constancy: for any $a\in \mathbb{R}$, $\varrho(a)=a$.
		\end{description}
	\end{Definition}
	
	\begin{Remark}
		 $\mathbf{B1}-\mathbf{B2}$ are well known and have been studied in detail in works on convex risk measures (see, for instance, F\"{o}llmer and Schied \cite{12}).  $\mathbf{B3}$ can be understood as a technical condition.
	\end{Remark}

	\begin{Remark}
		The two mappings in Definition~\ref{D31} and Definition~\ref{D32} divide the measurement of systemic risk into two steps, the deterministic function converts the uncertainty of systemic risk into certainty, and the convex  single-firm risk measure quantifies the risk simplified by the deterministic function.
	\end{Remark}

	

	We now introduce the definition of convex systemic risk measures in the variable-exponent Bochner-Lebesgue space $L^{p(\cdot)}$ via an axiomatic approach.

	\begin{Definition}\label{D33}
		A convex systemic risk measure is a functional 
		\begin{align*}
			\rho: L^{p(\cdot)}\ni f \mapsto \rho(f)\in\mathbb{R}\cup \{+\infty\}
		\end{align*}
		that satisfies the following properties:
		\begin{description}
			\item[C0] Surjectivity: $\rho(E)=\mathfrak{L}$ with $\mathfrak{L} = \mathbb{R}$ or $\mathfrak{L} = \mathbb{R}_{+}$. Here, $E$ is identified with the set of all constant mappings from $\Omega$ to $E$ (or, equivalently, deterministic $E$-valued payoffs);
			\item[C1] Monotonicity: for any $f,g\in L^{p(\cdot)}$, $f(\omega)\geq_{K} g(\omega)$ implies $\rho(f)\geq \rho(g)$;
			\item[C2] Preference consistency: If $\rho(f(\omega))\leq \rho(g(\omega))$ for all $\omega \in \Omega$, then $\rho(f)\leq \rho(g)$;
			\item[C3] Convexity: for any $f,g\in L^{p(\cdot)}$ and $\lambda\in[0,1]$, $\rho(\lambda f+(1-\lambda)g)\leq \lambda\rho(f)+(1-\lambda)\rho(g)$;
			\item[C4] Risk convexity: if $\rho(h(\omega))= \lambda\rho(f(\omega))+(1-\lambda)\rho(g(\omega))$ for a given scalar $\lambda\in[0,1]$ and for all $\omega \in \Omega$, then $\rho(h)\leq \lambda\rho(f)+(1-\lambda)\rho(g)$;
			\item [C5] There exists a constant $q\in [1,\infty]$ such that 
			\begin{align*}
				\big\{ f_{\rho}: \Omega \ni \omega \mapsto \rho(f(\omega)) \in \mathbb{R} \ \big| \ f \in L^{p(\cdot)} \big\} = L^{q}
			\end{align*}
			\begin{align*}
				\big(\textrm{respectively} \ \big\{ f_{\rho}: \Omega \ni \omega \mapsto \rho(f(\omega)) \in \mathbb{R}_{+} \ \big| \ f \in L^{p(\cdot)} \big\} = L^{q}_{+} \  \textrm{for} \  \mathfrak{L} = \mathbb{R}_{+}\big),
			\end{align*}
			where, for each fixed $\omega\in\Omega$, $\rho(f(\omega))$ is understood as the image of the constant mapping (or, sure payoff) 
			\begin{align*}
				\Omega\ni \omega^{\prime}\mapsto f(\omega)\in E
			\end{align*}
			under the functional $\rho$.
		\end{description}
		
	\end{Definition}
	
	\begin{Remark}
	 $\mathbf{C1}$ and $\mathbf{C3}$ can be interpreted in the same way as in the definition of single-firm risk models. 
	 $\mathbf{C2}$ means that if the economic risk $f(\omega)\in E$ is greater than the economic risk $g(\omega)\in E$ for almost all $\omega \in \Omega$, then the random economic risk $f\in L^{p(\cdot)}$ should be greater than the random economic risk of $g\in L^{p(\cdot)}$.
	 $\mathbf{C4}$ states that if the economic risk $h(\omega)$ is a convex combination of the risks of the economies $f(\omega)$ and $g(\omega)$ for all $\omega \in \Omega$, then the random economic risk of the economy $h \in L^{p(\cdot)}$ is at most the risk of the convex combination of the random economies $f, g \in L^{p(\cdot)}$.  $\mathbf{C5}$ is a technical requirement.  $\mathbf{C0}$ and $\mathbf{A0}$ of the corresponding functions are closely linked, and we will use these properties for our decomposition of the measurements in the following section; specifically, the condition $\phi(E)=\mathbb{R}=\rho(E)$ is needed.
		
	\end{Remark}

	\begin{Definition}\label{D34} 
		[1] A coherent deterministic function is a function  $\phi:E \rightarrow \mathbb{R}$ with $ \big\{ f_{\phi}: \Omega \ni \omega \mapsto \phi(f(\omega)) \in \mathbb{R} \ \big| \ f \in L^{p(\cdot)} \big\} \subseteq  L^{q}$ \big(respectively $ \big\{ f_{\phi}: \Omega \ni \omega \mapsto \phi(f(\omega)) \in \mathbb{R}_{+} \ \big| \ f \in L^{p(\cdot)} \big\} \subseteq  L^{q}_{+}$ for $\mathfrak{L} = \mathbb{R}_{+}$\big) for a certain $q\in [1,\infty]$ that satisfies $\mathbf{A1} - \mathbf{A2}$ and has the following properties:
		\begin{description}
			\item[A4] Positivite homogeneity: for any $x\in E$ and $t \in \mathbb{R}_{+}$, $ \phi(tx) = t \phi(x) $;
			\item[A5] Normalization: $\phi(z) = 1$.
		\end{description}
		[2]	A coherent single-firm risk measure is a function $\varrho:$ $L^{q} \rightarrow \mathbb{R}\cup \{+\infty\}$ that satisfies  $\mathbf{B1} - \mathbf{B2}$ and the following properties:
		\begin{description}
			\item[B4] Positivite homogeneity: for any $X\in L^{q}$ and $t \in \mathbb{R}_{+}$,  $\varrho(tX) = t\varrho(X)$;
			\item[B5] Normalization:  $\varrho(1) = 1$.
		\end{description}
		[3]	A coherent systemic risk measure is a function $\rho$: $L^{p(\cdot)}$ $\rightarrow$ $\mathbb{R}\cup \{+\infty\}$ with $\big\{ f_{\rho}: \Omega \ni \omega \mapsto \rho(f(\omega)) \in \mathbb{R} \ \big| \ f \in L^{p(\cdot)} \big\} \subseteq L^{q}$ \big(respectively $ \big\{ f_{\rho}: \Omega \ni \omega \mapsto \rho(f(\omega)) \in \mathbb{R}_{+} \ \big| \ f \in L^{p(\cdot)} \big\}  \subseteq  L^{q}_{+}$ for $\mathfrak{L} = \mathbb{R}_{+}$\big) for a certain $q\in [1,\infty]$ that satisfies  $\mathbf{C1} - \mathbf{C4}$  and the following properties:
		\begin{description}
			\item[C6] Positivite homogeneity: for any $f\in L^{p(\cdot)}$ and $t \in \mathbb{R}_{+}$, $\rho(tf) = t\rho(f)$;
			\item[C7] Normalization:  $\rho(z) = 1$.
		\end{description}
		
	\end{Definition}


	In the upcoming Section \ref{sec:4}, we will demonstrate how to compose a deterministic function with a single-firm risk measure to complete the construction of systemic risk measures under market volatility.

	\section{Systemic risk measures on $L^{p(\cdot)}$}
	\label{sec:4}
	
	Notably, a direct quantification of the convex systemic risk measure defined in Definition~\ref{D33} would reduce the financial system to a single portfolio-an innocuous simplification for non-systemic risks yet inappropriate for systemic ones. As Chen et al. \cite{8} emphasize, monetary risk measures should not be naively applied to an economy-wide aggregate; instead, they must capture inter-institutional interactions, because distinct financial institutions have different equity holders whose combined societal impact cannot be recovered by simple summation.

	Therefore, we decompose the quantification of systemic risk under market volatility using a deterministic function and a single-firm risk measure. First, a deterministic mapping dynamically normalizes systemic risk positions, reducing multi-factor systemic risk to a measurable variable driven by a single factor, thereby filtering out market volatility noise. Second, a single-firm risk quantification framework is introduced to assess the resulting univariate risk positions. By sequentially disentangling the systemic risk structure within a volatile environment, this approach preserves the traditional ability to capture common factor risk, while its modular design enhances cross-sector comparability and facilitates dynamic recalibration.

	In this section, we investigate systemic risk quantification on variable-exponent Bochner-Lebesgue space $L^{p(\cdot)}$, which provide a structural decomposition result that shows that each systemic risk measure in $L^{p(\cdot)}$ can be decomposed into a deterministic function and a single-firm risk measure. Furthermore, we show that any convex deterministic function and convex single-firm risk measure can be aggregated into a systemic risk measure in $L^{p(\cdot)}$. 
	
	\begin{Theorem}\label{T41}
		A functional $\rho$: $L^{p(\cdot)}$ $\rightarrow$ $\mathbb{R}\cup \{+\infty\}$ is a convex systemic risk measure if and only if there exist $q\in [1,\infty]$ such that 
		\begin{equation}\label{41}
			\rho(f) = (\varrho \circ \phi) (f), \quad f\in L^{p(\cdot)}
		\end{equation}
	where $\phi:E \rightarrow \mathbb{R}$ is a convex deterministic function  and $\varrho:$ $L^{q} \rightarrow \mathbb{R}\cup \{+\infty\}$ is a convex single-firm risk measure.  
	\end{Theorem}
	
	\noindent \textbf{Proof.}
	We first derive the ‘only-if' part. Suppose that $\rho$ is a convex systemic risk measure and define a function $\phi$ by
	\begin{equation}\label{42}
		\phi(x):=\rho(x)
	\end{equation}
	for any $x \in E$.
	As $\rho$ satisfies  $\mathbf{C0}$, it follows immediately from (\ref{42}) that $\phi$ satisfies  $\mathbf{A0}$. As $\rho$ satisfies $\mathbf{C1}$, for $f\geq_{K} g$, it follows $
	\phi\big( x \big)=\rho\big( x \big)\leq \rho\big( y \big)= \phi\big( y \big)
	$
	for any $x\in E$.  Thus, $\phi$ satisfies  $\mathbf{A1}$.
	Similarly, as $\rho$ satisfies the $\mathbf{C3}$, it follows that
	$
	\phi\big(\lambda x+(1-\lambda)y)\big)=\rho\big(\lambda x+(1-\lambda)y)\big)\leq \lambda\rho(x)+(1-\lambda)\rho(y)= \lambda\phi(x)+(1-\lambda)\phi(y)
	$
	for any $x,y\in E$ and $\lambda\in [0,1]$. Thus, $\phi$ satisfies $\mathbf{A2}$.   Moreover,  $\mathbf{A3}$ of $\phi$ can also be implied using  $\mathbf{C5}$.
	Therefore, the $\phi$ is a convex deterministic function.
	Next, we consider a functional $\varrho:$ $ \big\{ f_{\phi}: \Omega \ni \omega \mapsto \phi(f(\omega)) \in \mathbb{R} \ \big| \ f \in L^{p(\cdot)} \big\} \rightarrow \mathbb{R}\cup \{+\infty\}$, which is defined by
	\begin{equation}\label{43}
		\varrho(X):=\rho(f)
	\end{equation}
	with $ f\in L^{p(\cdot)} $ and $ \phi(f)=X $ where $ \phi $ is the convex deterministic function. 
	We now show that $\varrho$ is well defined. Suppose that $f,g \in L^{p(\cdot)}$ with $\phi(f)=\phi(g)$, we have
	$
	\rho(f(\omega))= \phi(f)(\omega) \geq \phi(g)(\omega) = \rho(g(\omega))
	$
	and
	$
	\rho(f(\omega))= \phi(f)(\omega) \leq \phi(g)(\omega) = \rho(g(\omega))
	$
	for all $\omega \in \Omega$. Thus, using  $\mathbf{C2}$ of $\rho$, we have $\rho(f(\omega)) = \rho(g(\omega))$ for almost all $\omega\in \Omega$ if $\phi(f)=\phi(g)$. Therefore, $\varrho$ is well defined. 
	Next, we show that the $\varrho$ defined above is a  convex single-firm risk measure. We suggest that $ \mathfrak{L} = \mathbb{R} $. With  $\mathbf{A3}$ of $ \phi $, it is easy to see that $ \big\{ f_{\phi}: \Omega \ni \omega \mapsto \phi(f(\omega)) \in \mathbb{R} \ \big| \ f \in L^{p(\cdot)} \big\} = L^{q}$ 
	for a constant $q\in [1,\infty]$. Suppose that $X,Y \in  \big\{ f_{\phi}: \Omega \ni \omega \mapsto \phi(f(\omega)) \in \mathbb{R} \ \big| \ f \in L^{p(\cdot)} \big\} $ 
	with $X \geq Y$. In this cases, there exists $f,g \in L^{p(\cdot)}$ such that $\phi(f)=X$, $\phi(g)=Y$. Then, we have
	$
	\rho(f(\omega))= \phi(f)(\omega) \geq \phi(g)(\omega) = \rho(g(\omega))
	$
	for all $\omega \in \Omega$. The case for $ \mathfrak{L} = \mathbb{R}_{+} $ can be proved similarly.   It follows again from  $\mathbf{C2}$ of $\rho$ that
	$
	\varrho(X)= \rho(f) \geq \rho(g) = \varrho(Y)
	$
	which implies that $\varrho$ satisfies  $\mathbf{B1}$. For $X, Y \in \phi(L^{p(\cdot)})$ and $\lambda\in [0,1]$, we consider $Z:= \lambda X + (1-\lambda) Y$. Suppose that $f, g, h \in L^{p(\cdot)}$ such that
	$
	\varrho(X)= \rho(f), \varrho(Y)= \rho(g), \varrho(Z)= \rho(h)
	$
	with $\phi(f)=X, \phi(g)=Y$ and $\phi(h)=Z$. Then,
	\begin{eqnarray*}
		\rho(h(\omega))&=&\phi(h)(\omega)\\&=&Z(\omega)\\
		&=&\lambda X(\omega)+(1-\lambda)Y(\omega)\\&=&\lambda\phi(f)(\omega)+(1-\lambda)\phi(g)(\omega)\\&=&\lambda\rho(f(\omega)) + (1-\lambda)\rho(g(\omega))
	\end{eqnarray*}
	for all $\omega \in \Omega$. Thus,  $\mathbf{C4}$ of $\rho$ yields
	$
	\varrho(Z)= \rho(h)\leq \lambda \rho(f) + (1-\lambda)\rho(g)= \lambda \varrho(X) + (1-\lambda)\varrho(Y).
	$
	Thus, $\varrho$ satisfies  $\mathbf{B2}$. From  $\mathbf{A0}$ of $\phi$ it follows that for any $a \in \mathbb{R}$, there exists $x \in E$ such that $\phi(x)=a$. Then, we have $\varrho(a)= \rho(x)$. Thus, $\rho(x) = \phi(x) = a$, and this implies that $\varrho(a)= a$ for any $a \in \mathbb{R}$, which means that $\varrho$ satisfies  $\mathbf{B3}$. Therefore, $\varrho$ is a  convex single-firm risk measure, and from (\ref{42}) and (\ref{43}), we have $\rho = \varrho\circ \phi$.
	
	Next, we derive the `if' part. We define a function $\rho$ by  
	\begin{equation}\label{431}
		\rho = \varrho\circ \phi
	\end{equation}
	where $\phi$ is a convex deterministic function and $\varrho$ is a  convex single-firm risk measure. 
	As $\varrho$ and $\phi$ are monotonic and convex, it is relatively simple to check that $\rho$ satisfies  $\mathbf{C1}$ and $\mathbf{C3}$. We now suppose that $f,g \in L^{p(\cdot)}$ satisfies
	$
	(\varrho\circ \phi)(f(\omega))= \rho (f(\omega)) \geq \rho (g(\omega)) = (\varrho\circ \phi)(g(\omega))
	$
	for all $\omega \in \Omega$. Then,  $\mathbf{B3}$ of $\varrho$ implies
	$
	\phi(f(\omega))\geq \phi(g(\omega))
	$
	for all $\omega \in \Omega$, which means that $\phi(f)\geq \phi(g)$. Therefore, using  $\mathbf{B1}$ of $\varrho$, we have
	$
	\rho(f)=(\varrho\circ \phi)(f) \geq (\varrho\circ \phi)(g)=\rho(g),
	$
	which indicates that $\rho$ satisfies  $\mathbf{C2}$. Next, we show that $\rho$ satisfies  $\mathbf{C4}$. In this case, we suppose that $f,g,h \in L^{p(\cdot)}$ and $\lambda \in [0,1]$ with
	$
	\rho(h(\omega))= \lambda \rho (f(\omega)) + (1-\lambda) \rho(g(\omega))
	$
	for all $\omega \in \Omega$. Therefore,
	$
	(\varrho\circ \phi)(h(\omega))= \lambda (\varrho\circ \phi)(f(\omega)) + (1-\lambda) (\varrho\circ \phi)(g(\omega))
	$
	for all $\omega \in \Omega$.  Then,  $\mathbf{B3}$ of $\varrho$ implies
	$
	\phi(h(\omega))= \lambda \phi(f(\omega)) + (1-\lambda) \phi(g(\omega))
	$
	for all $\omega \in \Omega$, which yields $\phi(h)= \lambda \phi(f) + (1-\lambda) \phi(g)$. Hence, using  $\mathbf{B2}$ of $\varrho$, we have
	$
	\rho(h) = (\varrho\circ \phi)(h) = \varrho\big( \lambda \phi(f) + (1-\lambda) \phi(g)  \big)  \leq \lambda (\varrho\circ \phi)(f) + (1-\lambda) (\varrho\circ \phi)(g)= \lambda \rho(f) + (1-\lambda) \rho(g).
	$
	Thus, $\rho$ satisfies  $\mathbf{C4}$. Moreover,  $\mathbf{C5}$ of $\rho$ is implied by $\mathbf{A3}$ and $\mathbf{B3}$.
	Now, we need only to show that $\rho$ satisfies  $\mathbf{C0}$. Using  $\mathbf{B3}$ of $\varrho$ and  $\mathbf{A0}$ of $\phi$, we have
	$
	\rho(E) = \varrho(\phi(E)) = \varrho(\mathbb{R}) = \mathbb{R},
	$
	which is just the   $\mathbf{C0}$ of $\rho$. Thus, $\rho$ defined in (\ref{431}) is a convex systemic risk measure.\qed\\
	
	\begin{Remark}
		Theorem~\ref{T41} not only offers a decomposition result for convex systemic risk measures in $L^{p(\cdot)}$ but also introduces a framework for addressing the systemic risk in markets characterized by uncertainty and volatility. Specifically, we first apply the convex deterministic function $\phi$ to transform the uncertainty of systemic risk into a certain quantity, and then we quantify this simplified risk using a convex single-firm risk measure.
		As a result, a regulator responsible for measuring systemic risk can develop a reasonable systemic risk measure by selecting an appropriate deterministic function and a suitable single-firm risk measure. The deterministic function should capture the relevant preferences concerning the uncertainty and volatility of the financial market.
	\end{Remark}

	The following theorem concerns coherent systemic risk measures on the variable-exponent Bochner-Lebesgue space $L^{p(\cdot)}$. The decomposition parallels that of Theorem~\ref{T41}, with the two component functions now provided by the coherent deterministic function and the coherent single-firm risk measure introduced in Definition~\ref{D34}.

	\begin{Theorem}\label{T43}
		A function $\rho$: $L^{p(\cdot)}$ $\rightarrow$ $\mathbb{R}\cup \{+\infty\}$ is a coherent systemic risk measure if and only if there exists a coherent deterministic function $\phi:E \rightarrow \mathbb{R}$ and a coherent single-firm risk measure $\varrho:$ $L^{q} \rightarrow \mathbb{R}\cup \{+\infty\}$ such that $\rho$ is the composition of $\varrho$ and $\phi$, i.e.
		\begin{equation}\label{44}
			\rho(f) = (\varrho \circ \phi) (f), \quad f\in L^{p(\cdot)}.
		\end{equation}
	\end{Theorem}
	
	\noindent \textbf{Proof.}
	We first show that, within the framework of Theorem~\ref{T41}, (\ref{44}) coincides with (\ref{41}). For a coherent deterministic function $\phi:E \rightarrow \mathbb{R}$,  $\mathbf{A0}$ is trivially satisfied. By the properties of the function $ \phi $, the function $ T_{\phi} $ defined in Lemma~\ref{L31} is given by $ T_{\phi}(a) = a $ for $ a \geq 0 $ and $ T_{\phi}(a) = -a \phi(-z) $ for $ a < 0 $ \big(respectively $T_{\phi}(a) = \phi(|a|z)= |a|  $ for $\mathfrak{L} = \mathbb{R}_{+}$\big). This is easy to imply $ \big\{ X_{T}: \Omega \ni \omega \mapsto T_{\phi}(X(\omega)) \in \mathbb{R} \ \big| \ X \in L^{q} \big\} = L^{q} $ \big(respectively $ \big\{ X_{T}: \Omega \ni \omega \mapsto T_{\phi}(X(\omega)) \in \mathbb{R}_{+} \ \big| \ f \in L^{p(\cdot)} \big\} =  L^{q}_{+}$ for $\mathfrak{L} = \mathbb{R}_{+}$\big) for a constant $q\in [1,\infty]$. Hence, for any $ X\in L^{q} $, there always exists $ Yz\in L^{p(\cdot)} $ such that $ X = T_{\phi}(Y) = \phi(Yz) $ which means $ L^{q} \subseteq \big\{ f_{\phi}: \Omega \ni \omega \mapsto \phi(f(\omega)) \in \mathbb{R} \ \big| \ f \in L^{p(\cdot)} \big\} $.  
	Simultaneously, since the definition of a deterministic function requires $\big\{ f_{\phi}: \Omega \ni \omega \mapsto \phi(f(\omega)) \in \mathbb{R} \ \big| \ f \in L^{p(\cdot)} \big\}  \subseteq L^{q} $,
	it follows that $ L^{q}  = \big\{ f_{\phi}: \Omega \ni \omega \mapsto \phi(f(\omega)) \in \mathbb{R} \ \big| \ f \in L^{p(\cdot)} \big\} $.
	Hence, under the framework of Theorem~\ref{T41}, (\ref{44}) is immediate. 
	
	For the `if' part, it remains to show that $ \rho $ satisfies  $\mathbf{C6}$ and $\mathbf{C7}$. It is easy to see  $\mathbf{C6}$ of $ \rho $ by  $\mathbf{A4}$ and $\mathbf{B4}$. With  $\mathbf{B3}$ of $ \varrho $ and  $\mathbf{A5}$ of $\phi$, we have $ \rho(z) = \varrho(\phi(z)) = \varrho(1) = 1 $, which means that  $\mathbf{C7}$ of $\rho$ is satisfied.
	
	Next, we derive the `only if' part, which remains to show that $ \phi $ satisfies  $\mathbf{A4}$ and $\mathbf{A5}$, while $ \varrho $ satisfies  $\mathbf{B4}$ and $\mathbf{B5}$. It is obvious to see that  $\mathbf{A4}$ and $\mathbf{A5}$ for $ \phi $ follow immediately from $ \rho $. Likewise,  $\mathbf{B4}$ for $ \varrho $ is inherited directly from $ \phi $ and $ \rho $. On the other hand, the proof of Theorem~\ref{T41}  reveals that $ \varrho $ satisfies  $\mathbf{B5}$.\qed \\

	The initial segment of the proof of Theorem~\ref{T43} directly invokes the conclusions of Theorem~\ref{T41}, since most of the axiomatic properties required for coherent systemic risk measures coincide with those of general systemic risk measures. Nevertheless, the imposition of positive homogeneity carries distinct implications, as detailed see Chen et al. \cite{8}.

	In the following section, we derive the dual representations of systemic risk measures in $L^{p(\cdot)}$ with the acceptance sets of $\phi$ and $\varrho$.
	
	\section{Dual representations of systemic risk measures in $L^{p(\cdot)}$}
	\label{sec:5}

	Although the preceding section achieved a structured characterization of systemic risk positions through the two-stage decomposition involving a deterministic function and a single-firm risk measure, the resulting construction remains purely forward looking. It prescribes how complex multi-institutional exposures are mapped into measurable variables, yet it furnishes no computable theoretical framework. In other words, knowing how to construct is insufficient to answer the central question of how to quantify. Consequently, the dual representationd of the systemic risk measured are needed.
	
	Dual representations confer several decisive advantages for risk quantification. By transforming the originally intractable primal optimization problem into a computable convex conjugate formulation, they render regulatory parameters such as liquidity quotas operationally implementable. In addition, the dual variables explicitly quantify the marginal contribution of each institution to aggregate risk. Simultaneously, embedding the risk measure within the convex analytic framework furnished by duality supplies a unified apparatus for subsequent sensitivity analysis, robustness testing, and model calibration.

	Before studying the dual representations of convex systemic risk measures in $L^{p(\cdot)}$, we first introduce the acceptance sets. Since every systemic risk measure $\rho$ can be decomposed into a convex deterministic function $\phi$ and a  convex single-firm risk measure $\varrho$, we next to define the acceptance sets of $\phi$ and $\varrho$ as follows:
	
	\begin{equation}\label{51}
		\mathcal{A}_{\varrho}:=\big\{(c,X)\in \mathbb{R}\times L^{q} \ \big| \  \varrho(X)\leq c \big\}
	\end{equation}
	and
	\begin{equation}\label{52}
		\mathcal{A}_{\phi}:=\big\{(Y,f)\in L^{q}\times L^{p(\cdot)} \ \big| \  \phi(f)\leq Y \big\}.\\
	\end{equation}
	
	We will see later that these acceptance sets can be used to provide a  representation result for systemic risk measures in $L^{p(\cdot)}$. The following properties  are required for the subsequent study.
	
	\begin{Definition}
		Let $M$ and $N$ be two ordered linear spaces. A set $A \subset M \times N$ satisfies l-monotonicity if $(m,n)\in A$, $u\in N$ and $n\geq u$ imply $(m,u)\in A$.  A set $A \subset M \times N$ satisfies b-monotonicity if $(m,n)\in A$, $v\in M$ and $v\geq m$ imply $(v,n)\in A$.
	\end{Definition}

	\begin{Proposition}\label{P1}
		Suppose that $\rho = \varrho\circ\phi$ is a systemic risk measure with deterministic function $\phi:E \rightarrow \mathbb{R}$ and a single-firm risk measure $\varrho:$ $L^{q} \rightarrow \mathbb{R}\cup \{+\infty\}$. The corresponding acceptance sets $\mathcal{A}_{\varrho}$ and $\mathcal{A}_{\phi}$ are defined by (\ref{51}) and (\ref{52}). Then, $\mathcal{A}_{\phi}$ and $\mathcal{A}_{\varrho}$ are convex sets and both of them satisfy the l-monotonicity and b-monotonicity.
	\end{Proposition}
	
	\noindent \textbf{Proof.}
	It is easy to check the above properties by the definitions of $\phi$ and $\varrho$. \qed\\
	
	The following proposition presents the primal representation of systemic risk measures in $L^{p(\cdot)}$ considering the acceptance sets. This result serves as the foundation for the representation of systemic risk measures in $L^{p(\cdot)}$.

	\begin{Proposition}\label{P2}
		Suppose that $\rho = \varrho\circ\phi$ is a systemic risk measure with convex deterministic function $\phi:E \rightarrow \mathbb{R}$ and a  convex single-firm risk measure $\varrho:$ $L^{q} \rightarrow \mathbb{R}\cup \{+\infty\}$. The corresponding acceptance sets $\mathcal{A}_{\varrho}$ and $\mathcal{A}_{\phi}$ are defined by (\ref{51}) and (\ref{52}). Then, for any $f\in L^{p(\cdot)}$,
		\begin{equation}\label{53}
			\rho(f) = \inf \big\{c\in \mathbb{R} \ \big| \  (c,X)\in \mathcal{A}_{\varrho}, (X,f)\in \mathcal{A}_{\phi} \big\}
		\end{equation}
		with $\inf \emptyset = \infty$.
	\end{Proposition}
	
	\noindent \textbf{Proof.}
	As $\rho = \varrho\circ\phi$, we have
	\begin{equation}\label{54}
		\rho(f) = \inf \big\{c\in \mathbb{R} \ \big| \   (\varrho\circ\phi)(f)\leq c \big\}.
	\end{equation}
	By the definition of $\mathcal{A}_{\varrho}$, we know that
	\begin{equation}\label{55}
		\varrho(X) = \inf \big\{c\in \mathbb{R} \ \big| \  (c,X)\in \mathcal{A}_{\varrho}  \big\}
	\end{equation}
	for all $X\in L^{q}$. Then, from (\ref{54}) and (\ref{55}),
	$
	\rho(f) = \inf \big\{c\in \mathbb{R} \ \big| \  (c,\phi(f))\in \mathcal{A}_{\varrho}\big\}.
	$
	It is relatively simple to check that
	$
	\big\{c\in \mathbb{R} \ \big| \   (c,\phi(f))\in \mathcal{A}_{\varrho}\big\} = \big\{c\in \mathbb{R} \ \big| \   (c,X)\in \mathcal{A}_{\varrho}, (X,f)\in \mathcal{A}_{\phi} \big\}.
	$
	Thus,
	$
	\rho(f) = \inf \big\{c\in \mathbb{R} \ \big| \   (c,X)\in \mathcal{A}_{\varrho}, (X,f)\in \mathcal{A}_{\phi} \big\}.
	$\qed\\
	
	Throughout this section, we denote by $L^{s}$ the dual space of $L^{q}$ with $ \frac{1}{q} + \frac{1}{s} =1 $.
	 With the help of Proposition~\ref{P2}, we introduce the main result of this section: the representation result of the convex systemic risk measures in $L^{p(\cdot)}$.

	\begin{Theorem}\label{T51}
		Suppose $\rho = \varrho\circ\phi$, where $ \varrho $
		 is a lower semicontinuous  convex single-firm risk measure and $ \phi $ is a continuous convex deterministic function. Then, for any $f\in L^{p(\cdot)}$, $\rho(f)$ has the following form
		\begin{equation}\label{56}
			\rho(f) = \sup_{(\widehat{Y},\widehat{f})\in \mathcal{P}} \Big\{\langle\widehat{f},f   \rangle  - \alpha (\widehat{Y},\widehat{f}) \Big\}
		\end{equation}
		where $\alpha: L^{s} \times (L^{p(\cdot)})^{\ast}$ $\rightarrow$ $\mathbb{R}\cup \{+\infty\}$ is defined by
		\begin{displaymath}
			\alpha (\widehat{Y},\widehat{f}):= \sup_{\substack{(c,X)\in \mathcal{A}_{\varrho}\\(Y,g)\in \mathcal{A}_{\phi}}} \Big\{-c - \langle \widehat{Y},(Y-X) \rangle + \langle\widehat{f},g  \rangle  \Big\}
		\end{displaymath}
		and
		$
		\mathcal{P}:= \big\{ (\widehat{Y},\widehat{f})\in L^{s} \times (L^{p(\cdot)})^{\ast} \ \big| \   \alpha (\widehat{Y},\widehat{f})< \infty \big\}.
		$
	\end{Theorem}
	
	\noindent \textbf{Proof.}
	By Proposition~\ref{P2}, we have
	$
	\rho(f) = \inf \big\{c\in \mathbb{R} \ \big| \   (c,X)\in \mathcal{A}_{\varrho}, (X,f)\in \mathcal{A}_{\phi} \big\}
	$
	for any $f\in L^{p(\cdot)}$. Furthermore, we can rewrite this formula as
	\begin{equation}\label{57}
		\rho(f) = \inf_{(c,X)\in \mathbb{R}\times L^{q}} \big\{c + I_{\mathcal{A}_{\varrho}}(c,X) + I_{\mathcal{A}_{\phi}}(X,f) \big\}
	\end{equation}
	where the indicator function of  a set $A \in \mathcal{X}\times \mathcal{Y}$ is defined by
	\begin{displaymath}
		I_{A}(x,y):=\left\{ \begin{array}{ll}
			0, & (x,y)\in \mathcal{X}\times \mathcal{Y};\\
			\infty, & \textrm{otherwise.}
		\end{array} \right.
	\end{displaymath}
	From Proposition~\ref{P1}, we know that $\mathcal{A}_{\varrho}$ and $\mathcal{A}_{\phi}$ are convex sets. Thus,
	\begin{displaymath}
		I_{\mathcal{A}_{\varrho}}^{\prime}(\widehat{c},\widehat{X})= \sup_{(\overline{c},\overline{X})\in \mathcal{A}_{\varrho}} \big\{ \widehat{c} \overline{c} + \langle \widehat{X}, \overline{X}\rangle  \big\},\quad  \widehat{c}\in\mathbb{R}, \widehat{X}\in L^{s}
	\end{displaymath}
	and
	\begin{displaymath}
		I_{\mathcal{A}_{\phi}}^{\prime}(\widehat{Y},\widehat{f})= \sup_{(\overline{Y},\overline{f})\in \mathcal{A}_{\phi}} \big\{ \langle \widehat{Y},\overline{Y}\rangle + \langle \widehat{f}, \overline{f}\rangle   \big\},\quad  \widehat{Y}\in L^{s}, \widehat{f}\in (L^{p(\cdot)})^{\ast}.
	\end{displaymath}
	Next, because $\varrho$ is lower semicontinuous, it follows that $\mathcal{A}_{\varrho}$ is closed.  Thus, by the duality theorem for conjugate functions, we have
	\begin{eqnarray*}
		I_{\mathcal{A}_{\varrho}}(c,X) &=& I_{\mathcal{A}_{\varrho}}^{\prime\prime}(c,X)\\&=&\sup_{(\widehat{c},\widehat{X})\in \mathbb{R}\times L^{s}}\big\{ \widehat{c}c + \langle \widehat{X},X \rangle - I_{\mathcal{A}_{\varrho}}^{\prime}(\widehat{c},\widehat{X}) \big\}\\
		&=& \sup_{(\widehat{c},\widehat{X})\in \mathbb{R}\times L^{s}}\Big\{ \widehat{c}c + \langle \widehat{X},X \rangle - \sup_{(\overline{c},\overline{X})\in \mathcal{A}_{\varrho}} \big\{ \widehat{c} \overline{c} + \langle \widehat{X}, \overline{X}\rangle  \big\}\Big\}.
	\end{eqnarray*}
	Similarly, we have
	\begin{eqnarray*}
		I_{\mathcal{A}_{\phi}}(X,f) &=& I_{\mathcal{A}_{\phi}}^{\prime\prime}(X,f)\\&=&\sup_{(\widehat{Y},\widehat{f})\in L^{s}\times (L^{p(\cdot)})^{\ast}}\big\{  \langle \widehat{Y},X\rangle + \langle \widehat{f},f \rangle  - I_{\mathcal{A}_{\phi}}^{\prime}(\widehat{Y},\widehat{f}) \big\}\\
		&=& \sup_{(\widehat{Y},\widehat{f})\in L^{s}\times (L^{p(\cdot)})^{\ast}}\Big\{ \langle \widehat{Y},X\rangle + \langle \widehat{f},f \rangle  - \sup_{(\overline{Y},\overline{f})\in \mathcal{A}_{\phi}} \big\{ \langle \widehat{Y},\overline{Y}\rangle + \langle \widehat{f}, \overline{f}\rangle   \big\}\Big\}.
	\end{eqnarray*}
	Thus, we know that
	\begin{eqnarray*}
		\rho(f)&=& \inf_{(c,X)\in \mathbb{R}\times L^{q}} \big\{c + I_{\mathcal{A}_{\varrho}}(c,X) + I_{\mathcal{A}_{\phi}}(X,f) \big\}\\
		&=& \inf_{(c,X)\in \mathbb{R}\times L^{q}} \sup_{\substack {(\widehat{c},\widehat{X})\in \mathbb{R}\times L^{s}\\(\widehat{Y},\widehat{f})\in L^{s}\times (L^{p(\cdot)})^{\ast}}}\Big\{ c(1+\widehat{c}) +  \langle \widehat{X}+ \widehat{Y}, X\rangle + \langle \widehat{f},f \rangle  -\\
		&& I_{\mathcal{A}_{\varrho}}^{\prime}(\widehat{c},\widehat{X})- I_{\mathcal{A}_{\phi}}^{\prime}(\widehat{Y},\widehat{f})  \Big\}.
	\end{eqnarray*}
	By Theorem 7 of Rockafellar \cite{20}, because of the  lower semicontinuous of $\varrho$ and the continuity of $\phi$, we can interchange the supremum and the infimum above, i.e.,
	\begin{eqnarray*}
		\rho(f) &=& \sup_{\substack {(\widehat{c},\widehat{X})\in \mathbb{R}\times L^{s}\\(\widehat{Y},\widehat{f})\in L^{s}\times (L^{p(\cdot)})^{\ast}}}\inf_{(c,X)\in \mathbb{R}\times L^{q}}\Big\{ c(1+\widehat{c}) +  \langle \widehat{X}+ \widehat{Y}, X\rangle + \langle \widehat{f},f \rangle  -\\
		&& \qquad     I_{\mathcal{A}_{\varrho}}^{\prime}(\widehat{c},\widehat{X})- I_{\mathcal{A}_{\phi}}^{\prime}(\widehat{Y},\widehat{f})  \Big\}\\
		&=& \sup_{(\widehat{Y},\widehat{f})\in L^{s}\times (L^{p(\cdot)})^{\ast}} \Big\{  \langle \widehat{f},f \rangle  - \sup_{\substack {(\overline{c},\overline{X})\in \mathcal{A}_{\varrho}\\(\overline{Y},\overline{f})\in \mathcal{A}_{\phi}}} \big\{  -\overline{c}-  \langle \widehat{Y}, \overline{Y}-\overline{X}\rangle +   \langle \widehat{f},\overline{f} \rangle  \big\}  \Big\}.
	\end{eqnarray*}
	With $ \alpha (\widehat{Y},\widehat{f})$ being defined by
	\begin{eqnarray*}
		\alpha (\widehat{Y},\widehat{f}):&=&  \sup_{\substack {(\overline{c},\overline{X})\in \mathcal{A}_{\varrho}\\(\overline{Y},\overline{f})\in \mathcal{A}_{\phi}}} \big\{  -\overline{c}-  \langle \widehat{Y}, \overline{Y}-\overline{X}\rangle +   \langle \widehat{f},\overline{f} \rangle  \big\}\\
		&=&\sup_{\substack{(c,X)\in \mathcal{A}_{\varrho}\\(Y,g)\in \mathcal{A}_{\phi}}} \Big\{-c - \langle \widehat{Y},Y-X \rangle + \langle\widehat{f},g  \rangle  \Big\}
	\end{eqnarray*}
	and
	\begin{displaymath}
		\mathcal{P}:= \big\{ (\widehat{Y},\widehat{f})\in L^{s} \times (L^{p(\cdot)})^{\ast} \ \big| \  \alpha (\widehat{Y},\widehat{f})< \infty \big\},
	\end{displaymath}
	it immediately follows that
	\begin{displaymath}
		\rho(f) = \sup_{(\widehat{Y},\widehat{f})\in \mathcal{P}} \Big\{\langle\widehat{f},f   \rangle  - \alpha (\widehat{Y},\widehat{f}) \Big\}.
	\end{displaymath}
	\qed


	To derive the dual representation of a coherent systemic risk measure, we again follow the approach used for the systemic risk measure in Theorem~\ref{T51} and consider the acceptance sets of the coherent deterministic function $\phi$ and the coherent single-firm risk measure $ \varrho $, i.e.
	\begin{equation}\label{58}
		\overline{\mathcal{A}}_{\varrho}:=\big\{(b,\widehat{X})\in \mathbb{R}\times L^{s} \ \big| \   cb - \langle\widehat{X},X \rangle \geq 0  \textrm{ for any } (c,X)\in \mathcal{A}_{\varrho} \big\}
	\end{equation}
	and
	\begin{equation}\label{59}
		\overline{\mathcal{A}}_{\phi}:=\big\{(\widehat{Y},\widehat{f})\in L^{s}\times (L^{p(\cdot)})^{\ast} \ \big| \   \langle\widehat{Y},Y \rangle - \langle\widehat{f},f \rangle  \geq 0 \textrm{ for any } (Y,f)\in \mathcal{A}_{\phi}   \big\}.\\
	\end{equation}
	
	Indeed, the $ \overline{\mathcal{A}}_{\varrho} $ and $ \overline{\mathcal{A}}_{\phi} $ can be regarded as the dual cones of the $\mathcal{A}_{\varrho} $ and $ \mathcal{A}_{\phi} $.

	\begin{Theorem}\label{T52}
			Suppose $\rho = \varrho\circ\phi$, where $ \varrho $
		is a lower semicontinuous coherent single-firm risk measure and $ \phi $ is a continuous coherent deterministic function. Then, for any $f\in L^{p(\cdot)}$, $\rho(f)$ has the following form
		\begin{equation}\label{511}
			\rho(f) = \sup_{(\widehat{Y},\widehat{f})\in\overline{\mathcal{P}}} \Big\{\langle\widehat{f},f   \rangle  - \alpha (\widehat{Y},\widehat{f}) \Big\} 
		\end{equation}
		with $\alpha (\widehat{Y},\widehat{f}) = 0$ and $\overline{\mathcal{P}}$ is defined by
		\begin{displaymath}
			\overline{\mathcal{P}}:= \big\{ (\widehat{Y},\widehat{f})\in L^{s} \times (L^{p(\cdot)})^{\ast} \ \big| \   (1, \widehat{Y}) \in \overline{\mathcal{A}}_{\varrho}, (\widehat{Y},\widehat{f})\in \overline{\mathcal{A}}_{\phi}  \big\}.
		\end{displaymath}
	\end{Theorem}
	\noindent \textbf{Proof.}
	In light of Theorem~\ref{T51},
	$ \rho(f) $ admits the following representation,
	\[
	\rho(f) = \sup_{(\widehat{Y},\widehat{f})\in L^{s} \times (L^{p(\cdot)})^{\ast}} \Big\{\langle\widehat{f},f   \rangle  - \alpha (\widehat{Y},\widehat{f}) \Big\}
	\]
	with 
	\[
	\alpha (\widehat{Y},\widehat{f})= \sup_{\substack{(c,X)\in \mathcal{A}_{\varrho}\\(Y,g)\in \mathcal{A}_{\phi}}} \Big\{-c - \langle \widehat{Y},Y-X \rangle + \langle\widehat{f},g  \rangle  \Big\}, \quad (\widehat{Y},\widehat{f})\in L^{s} \times (L^{p(\cdot)})^{\ast}.
	\]
	For $ (1, \widehat{Y}) \in \overline{\mathcal{A}}_{\varrho}$ and $ (\widehat{Y},\widehat{f})\in \overline{\mathcal{A}}_{\phi} $, it is easy to check that $ -c - \langle \widehat{Y},Y-X \rangle + \langle\widehat{f},g  \rangle  \leq 0 $, which means $ \alpha (\widehat{Y},\widehat{f}) = 0 $ for $ (\widehat{Y},\widehat{f})\in \overline{\mathcal{P}} $. Thus, we get representation (\ref{511}). \qed

	\begin{Remark}
		Once the duality theory of the variable-exponent Bochner-Lebesgue space $L^{p(\cdot)}$ is coupled with the systemic risk measure developed in the preceding section, the stochastic fluctuations of the exponent $p(\cdot)$ become an instantaneous projection of market sentiment and local volatility. Theorem~\ref{T51} and Theorem~\ref{T52} deliver dual representations of systemic risk measures on $L^{p(\cdot)}$ that eliminate fixed sensitivity assumptions and offer regulators a rigorous analytical framework. In extreme market episodes or low-frequency and high-loss events, these representations yield dynamically adaptive policy instruments, such as capital surcharges, liquidity buffers, or position limits, that adjust in real time to current market volatility. Consequently, this framework significantly enhances the accuracy, robustness, and implementability of systemic risk quantification under pronounced market turbulence.
	\end{Remark}

	\section{Examples}
	\label{sec:6}
	
	The objective of this section is to apply the previously introduced systemic risk measures to practical scenarios. For the deterministic function, we select the Bochner integral as the aggregation operator on $L^{p(\cdot)}$ to normalize systemic risk under market volatility. For the single-firm risk measure, we illustrate several widely adopted specifications from recent literature. The aggregation framework developed in Section~\ref{sec:4} then synthesizes the deterministic function and the single-firm risk measures into a unified systemic risk measure.
	
	In the following examples, we set $ E = L^{q} $. Therefore, an element in $L^{p(\cdot)}$ is a function $f:\Omega\rightarrow L^{q}$ that satisfies the properties described in Section~\ref{sec:2}. Notably, in this case, the element in $L^{p(\cdot)}$ can be characterized as $ f: \omega_{1} \rightarrow k(\omega_{1}, \cdot) $, where $ k: \Omega \times \Omega \rightarrow \mathbb{R} $.
	
	\begin{Example}
		Let $ E = L^{q} $, then we can define the deterministic function by Bochner integral
		\begin{equation}\label{61}
			\phi(f) = \int_{\Omega} k( \cdot, \omega_{2}) d\mu.
		\end{equation}
		Consider the distortion entropic risk measure $\varrho_{d}$ introduced by Tsanakas and Desli \cite{31}, i.e. 
		\begin{equation}\label{62}
			\varrho_{d}(X) = \frac{1}{d} \log Q_{h}(\exp^{sX}),\qquad d>0
		\end{equation}
		where the distorted expection $ Q_{h} $ is defined by
		\[
		Q_{h}(X):= \int_{-\infty}^{0}\big[ h(\mu(X>x)) -1 \big]dx + \int_{0}^{\infty} h(\mu(X>x))dx
		\]
		with $ h : [0, 1] \rightarrow [0, 1] $ being non-decreasing function with $h(0) = 0$ and $h(1) = 1$.
		Hence, we can use the deterministic function (\ref{61}) and distortion entropic risk measure to define distortion entropic systemic risk measure by 
		\[
		\rho_{h, d}(f) = \frac{1}{d} \log Q_{h}(\exp^{d\int_{\Omega} k(\cdot, \omega_{2}) d\mu}),\qquad d>0.
		\]
	\end{Example}

	\begin{Example}
		Let $ E = L^{q} $, then we can define the deterministic function by Bochner integral
		\begin{equation}\label{63}
				\phi(f) = \int_{\Omega} k(\cdot, \omega_{2}) d\mu.
		\end{equation}
		We consider the expected shortfall $\varrho_{e}$ disscussed by Acharga et al. \cite{1}, i.e.
		\begin{equation}\label{64}
			\varrho_{e}(X) = ES_{e}(X) = \sup_{\mathbb{P}\in \mathcal{H}_{e} } E_{\mathbb{P}} (X)
		\end{equation}
		where 
		\[
		\mathcal{H}_{e} = \Big\{ \mathbb{P}< < \mu  \ \big| \   \frac{d\mathbb{P}}{d\mu} \leq  \frac{1}{1-e}    \Big\}.
		\]
		Hence, we can use the deterministic function (\ref{63}) and expected shortfall (\ref{64}) to define expected shortfall systemic risk measure by 
		\[
		\rho_{e}(f) = \sup_{\mathbb{P}\in \mathcal{H}_{e} } E_{\mathbb{P}}\Big(\int_{\Omega} k(\cdot, \omega_{2}) d\mu\Big).
		\]
	\end{Example}

	\begin{Example}
		Let $ E = L^{q} $, we define the following deterministic function
		\begin{equation}\label{65}
			\phi(f) = \Big(\int_{\Omega} k(\cdot, \omega_{2}) d\mu\Big)^{+},
		\end{equation}
		with $ (X)^{+} := \max\{X, 0\} $.
		The loss-based risk measures were first investigated by Cont et al. \cite{32}. Next, we consider the following single-firm risk measure,
		\begin{equation}\label{66}
			\varrho_{l}(X) =  E_{\mathbb{Q}} (X)
		\end{equation}
		where $ \mathbb{Q} $ is a given probability with $ \mathbb{Q} <<\mu $. 
		Hence, we can use the deterministic function (\ref{65}) and single-firm risk measure (\ref{66}) to define the loss-based systemic risk measure by 
		\[
		\rho_{l}(f) =  E_{\mathbb{Q}} \Big[\Big(\int_{\Omega} k(\cdot, \omega_{2}) d\mu\Big)^{+}\Big].
		\]	 
	\end{Example}

	\section{Conclusion}
	
	Systemic risk is the type of risk with the greatest threat to financial stability, and its accurate quantification is essential for economic security and public trust. However, existing systemic risk measures are not adequately adapted to market volatility.
	
	In this paper, we adopt the variable-exponent Bochner-Lebesgue space $L^{p(\cdot)}$ to model systemic risk positions under market volatility, where the random exponent $p(\cdot)$ is used to characterize the fluctuations in the financial market. Our principal contribution is the development of systemic risk measures on $L^{p(\cdot)}$ together with their dual representations.
	
	Our framework is built upon two functions. First, a deterministic function dynamically normalizes systemic risk positions, converting multi-factor systemic risk into a measurable variable driven by a single factor, thereby filtering out market volatility noise. Second, a single-firm risk measure is introduced to evaluate the resulting univariate risk exposure.
	Next, an interesting and relevant problem is the dual representation of the systemic risk measure. By leveraging the duality of the space $L^{p(\cdot)}$, we derive a complete dual representation of the proposed systemic risk measures. Moreover, in the final section, we provide several concrete examples of the resulting systemic risk measures.
	
	For potential future research, a natural direction is to apply systemic risk measures on $L^{p(\cdot)}$ to some specific risk quantification problems.
	It is also interesting to investigate other risk measures on $L^{p(\cdot)}$.


	%
	%

	

\end{document}